\documentclass[twocolumn,traditabstract]{aa} 
\pdfoutput=0

\usepackage{amsmath,amssymb,latexsym}
\usepackage{natbib}
\usepackage{graphicx}
\usepackage{color}

\definecolor{lila}{rgb}{0.5,0,1}
\definecolor{grau}{rgb}{0.5,0.5,0.5}

\newcommand{\ME}{M_{\oplus}}

\newcommand{\RJ}{R_{\rm J}}
\newcommand{\Mcore}{M_{\rm core}}
\newcommand{\Zatm}{Z_{\rm atm}}
\newcommand{\Ptrans}{P_{\rm trans}}
\newcommand{\ToFthree}{ToF$\,$3 }
\newcommand{\ToFfour}{ToF$\,$4 }

\begin{document} 

\title{Low- and high-order gravitational harmonics of rigidly rotating Jupiter}


\author{N.~Nettelmann}
\institute{Universit\"at Rostock, Institut f\"ur Physik, 18051 Rostock, Germany} 
\abstract{
The Juno Orbiter has provided improved estimates of the even gravitational harmonics $J_2$ 
to $J_8$ of Jupiter. To compute  higher-order moments, new methods such as the Concentric Maclaurin Spheroids 
(CMS) method have been developed  which surpass the so far commonly used Theory of Figures (ToF) method in accuracy.
This progress rises the question whether ToF can still provide a useful service for deriving the internal 
structure of giant planets in the Solar system. 
In this paper, I apply both the ToF and the CMS method to compare results for polytropic Jupiter and for 
the physical equation of state H/He-REOS.3 based models.
An accuracy in the computed values of $J_2$ and $J_4$ of 0.1\% is found to be sufficient in order to
obtain the core mass safely within $0.5\:\ME$ numerical accuracy and the atmospheric metallicity within 
about 0.0004. ToF to 4th order provides that accuracy,  while ToF to 3rd order does not for $J_4$.
Furthermore, I find that the assumption of rigid rotation yields $J_6$ and $J_8$ values in agreement with 
the current Juno estimates, and that higher order terms ($J_{10}$ to $J_{18}$) deviate by about $10\%$ from 
predictions by polytropic models. 
This work suggests that \ToFfour can still be applied to infer the deep internal structure, and that the
zonal winds on Jupiter reach less deep than 0.9 $\RJ$.
}

\titlerunning{ToF and CMS}
\maketitle 

\section{Introduction}

The Theory of Figures \citep{ZT78} to third or fourth order, hereafter labeled respectively ToF$\,$3 and ToF$\,4$, 
is commonly used to compute the gravity field of the gas giant planets in the Solar system 
(e.g.,~\citealp{SauGui04,Helled11,LC12,Nettelmann12,HellGui13,Miguel16}). While a theory to $n$-th order 
allows to compute the gravity field in terms of the gravitational harmonics up to $J_{2n}$ only, Jupiter's gravity 
field before the current Juno mission was also measured up to $J_6$ only. Thus, 3rd or 4th order theories seemed 
sufficient. Moreover, it is the low-order harmonics $J_2$ and $J_4$ which are particularly sensitive to the 
internal density distribution; they allow to derive the interior structure parameters core mass and envelope 
metallicity. Hence for a long time, the observational data 
of gravity field theories (e.g., ToF), and planet interior parameters of interest formed a closed system. 

This convenient situation has changed with the arrival of the Juno spacecraft at Jupiter. Juno's sensitivity limit allows
to measure the rigid-rotation contribution to the gravitational harmonics up to $J_{14}$ \citep{Kaspi10}. 
High-order moments yield clues on the properties of the zonal winds as the flows influence the density distribution 
which in turn is the source function of the gravitational potential. Differential rotation due to zonal flows is 
predicted to entirely dominate the $J_{2n}$ for $n\geq 14$, while to be within a factor of 10 of the prediction 
for a rigidly rotating planet for $J_8$--$J_{12}$ \citep{Hubbard99,Kaspi10,CaoStev17}. Since the wind 
contribution $\Delta J_{2n}^{\rm wind}$ is obtained by subtracting the theoretical values for a rigidly rotating 
planet from the observed ones ($J_{2n}^{\rm obs} = J_{2n}^{\rm rigid}$ + $\Delta J_{2n}^{\rm wind}$), it is also 
important to have accurate knowledge of the rigid-rotation contribution. 

For that purpose, \citet{Hubbard13}, hereafter H13, developed the Concentric Maclaurin Spheroids (CMS) method. 
This method yields demonstratively 
good agreement with the exact Bessel solution for an $n=1$ polytrope model of Jupiter. Deviations have been found 
to be about $5\times 10^{-5}$ in $J_2$ to $2\times 10^{-4}$ in $J_{20}$ \citep{WH16}, or to be of order $2\times 10^{-3}$ 
\citep{CaoStev17}. However, comparison of the exact Bessel solution to the \ToFthree results (H13;~\citealp{WH16suppl}) 
have led to the conclusion of \ToFthree being of insufficient accuracy for modeling Jupiter \citep{WH16suppl}. 
This rises the question of what accuracy in the low-order moments is desired for inferring Jupiter's internal 
density distribution, and which methods can provide that.

\begin{figure*}[t]
\centering
\includegraphics[angle=0,width=0.86\textwidth]{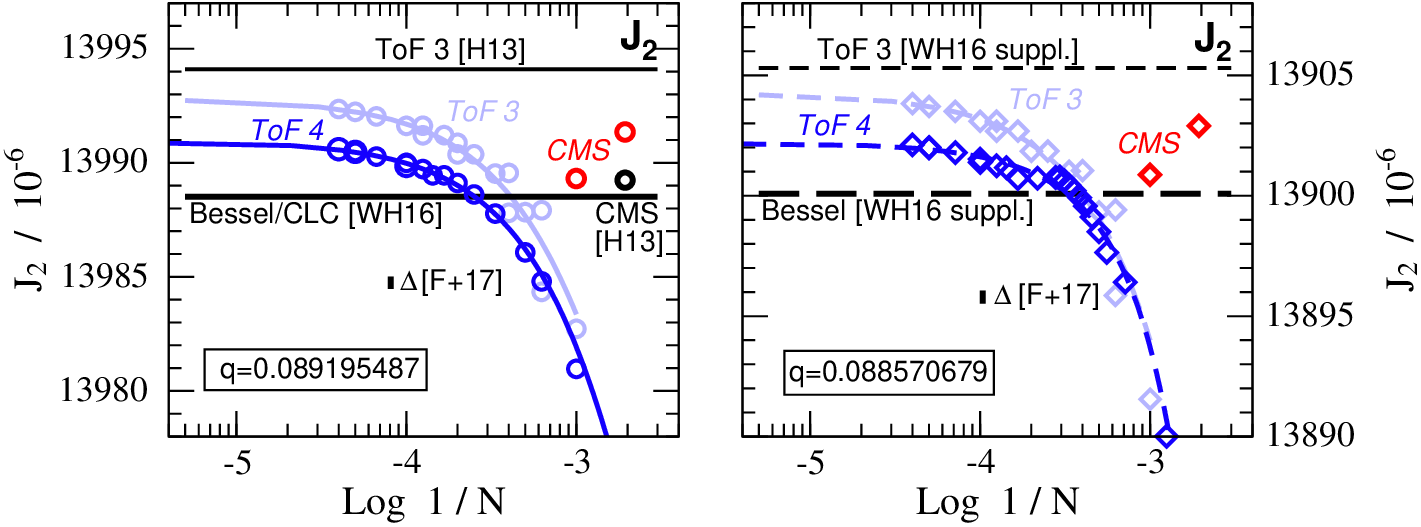}
\caption{Resulting $J_2$ values of polytropic models of rigidly rotating Jupiter for $q=0.089195487$ (\emph{left})
as in H13 and \cite{WH16}, and for $q=0.088570679$ (\emph{right}) as in \cite{WH16suppl}. \emph{Blue symbols:} 
using ToF to $4^{\rm th}$ order, \emph{blue lines:} respective fit curves, \emph{light blue}: same as blue but 
using \ToFthree,  \emph{red symbols}: using CMS method. Reference values are in \emph{black}; \emph{black circle}: 
CMS results of H13 for $N=512$, \emph{horizontal black lines}: unknown value of $N$, in particular: 
\emph{thick black lines}: Bessel/CLC results of \citet{WH16}, \emph{thin black lines:} \ToFthree results of H13 and 
\citet{WH16suppl}. The \emph{vertical black arrow} shows the current Juno uncertainty of $J_2$ \citep{Folkner17}, 
here arbitrarily placed mid $x$-axis. The $x$-axis is number of radial grid points $N$.}
\label{fig:polyJ2}
\end{figure*}

\begin{figure*}[htb]
\centering
\includegraphics[angle=0,width=0.86\textwidth]{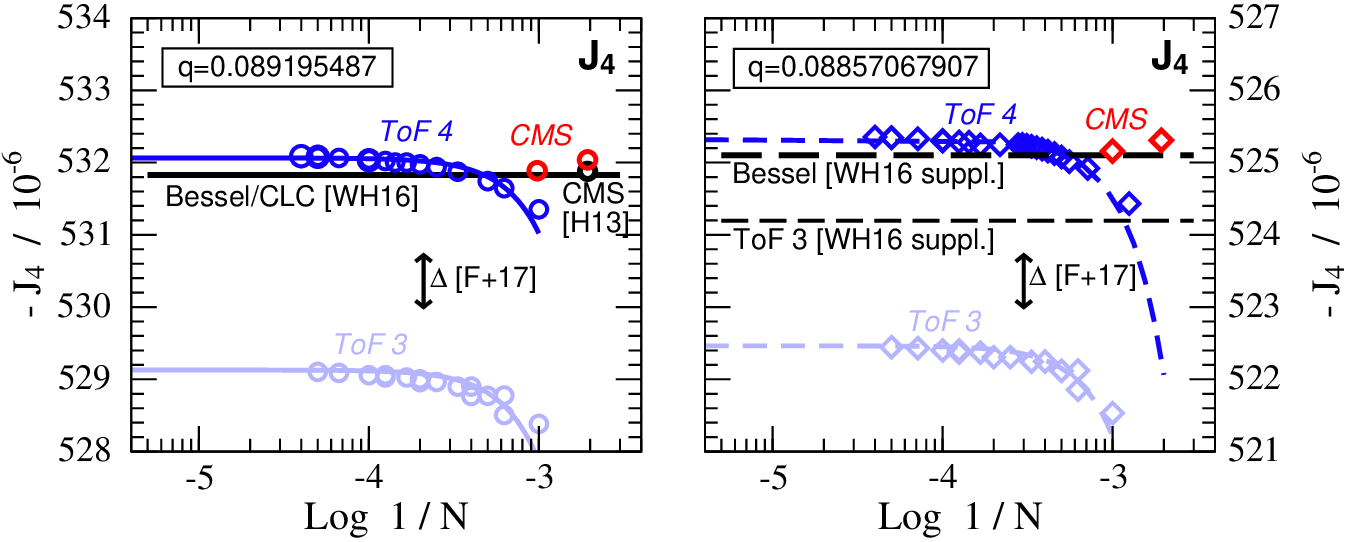}
\caption{Same as Figure \ref{fig:polyJ2} but for $J_4$. The \ToFthree result of \citet{Hubbard13} exceeds the shown 
range of $J_4$ values.}
\label{fig:polyJ4}
\end{figure*}

The classical view of a Jupiter-like gas giant is that of 
a well-defined core embedded into an H/He-rich envelope, in which case one can ask for the mass of the core 
and the heavy element mass fraction of the envelope and use the measured planet radius and $J_2$ value to 
determine both \citep{Militzer08}. However, recent gas giant formation models challenge that simple picture. 
Simultaneous accretion of gas and planetesimals might naturally lead to a gas-enriched, i.e.~diluted core 
\citep{Venturini16}, where the metallicity is predicted to decrease outward as a result of the slow convective 
timescale compared to the accretion rate during formation \citep{HelledStev17}; it may remain permanent due to the 
inhibition of convection once a compositional gradient has established \citep{Vazan16}. 
Diluted cores have been found to enhance the predicted atmospheric metallicity of Jupiter models \citep{ForNett10}.

In this paper, the uncertainty in the computed values for the low-order harmonics $J_2$, $J_4$, $J_6$ 
due to application of ToF$\,$3 and ToF$\,$4 is estimated by the using $n=1$ polytrope model (Section \ref{sec:poly});
the corresponding uncertainty in the derived core mass and envelope metallicity of Jupiter is estimated in
Section \ref{sec:modelsJ2}. Finding this uncertainty to be small for ToF$\,$4, I use this method in Section 
\ref{sec:models3L} to compute physical EOS based Jupiter models that are designed to match the low-order harmonics 
$J_2$ and $J_4$ from Juno's first two low-periapse polar orbits around Jupiter \citep{Folkner17}. Models are 
presented both for solid cores and for diluted cores, as well as for deep zonal wind corrections as proposed by 
\citet{CaoStev17}. For some of the Jupiter models I compute the high-order moments using the CMS method 
(Section \ref{sec:highJs}), thereby providing the first prediction of the high-order $J_{2n}$ values for a model 
of adiabatic, rigidly rotating Jupiter that matches the measured low-order moments. 
Conclusions are in Section \ref{sec:conclusion}. In Appendix A my implementation of the CMS method is
validated for the linear density case, while in Appendix B the ToF coefficients are provided up to 4th order.

\section{Polytropic models}\label{sec:poly}

In this Section, $n=1$ polytropic models are computed for $GM_{\rm J}=12.6686536\times 10^{16}$ $\rm m^3/s^2$, 
equatorial radius $R_{\rm eq}=\RJ=71492$~km, and for two different rotation rates as represented by $q=0.0891954870$  
\citep{WH16} and $q=0.08857067907$ \citep{WH16suppl}, where $q=\omega^2 R_{\rm eq}^3 / GM$. 
I apply ToF to 3rd and 4th order as well as the CMS method. For both methods an iterative procedure 
is required to ensure the total mass is conserved, and that for the thus specified value of $K$  in the 
polytropic relation $P=K\rho^2$ hydrostatic balance holds. 

\begin{figure}[htb]
\centering
\includegraphics[angle=0,width=0.45\textwidth]{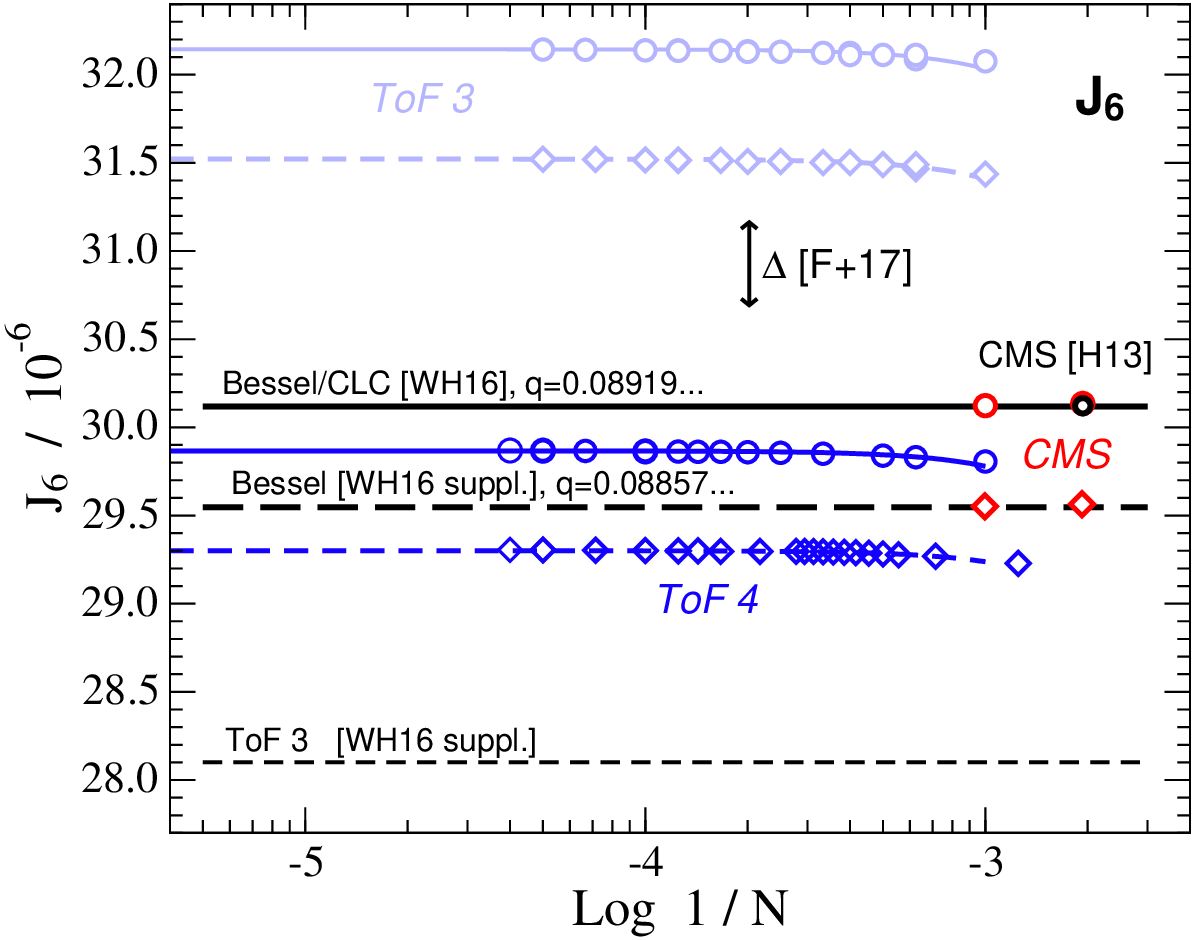}
\caption{Same as Figures \ref{fig:polyJ2} and \ref{fig:polyJ4} but for $J_6$ and results for different $q$-values 
merged into a single panel, distinguished by respectively \emph{solid lines/circles} and \emph{dashed lines/diamonds}. 
The \ToFthree result of \citet{Hubbard13} exceeds the shown range of $J_6$ values.}
\label{fig:polyJ6}
\end{figure}

 
With ToF I calculate the density at grid 
point $i$ using $\rho_i=\sqrt{P_i/K}$, while with CMS method $\rho_i=\sqrt{ 0.5(P_i+P_{i+1})/K}$ (H13) 
except for $i=0$ where $\rho_0=P_0=0$ in their respective units.
As I find the dependence on the number of radial grid points, $N$ to be strong, I plot the resulting 
values of $J_2$ (Fig.~\ref{fig:polyJ2}), $J_4$ (Fig.~\ref{fig:polyJ4}), and $J_6$ (Fig.~\ref{fig:polyJ6})
against $N$.

Compared to the exact Bessel solution \citep{WH16}, CMS method performs best and \ToFthree worst. In particular,
\ToFthree underestimates $|J_4/10^{-6}|$ by 1--2 times (depending on the implementation) the pre-Juno $1\sigma$ 
error bar of $\sim 2$ \citep{Miguel16}, and  therefore predicts a higher atmospheric metallicity  for Jupiter 
than \ToFfour does \citep{Nettelmann12}.

ToF$\,4$, on the other hand, performs much better: the difference in $J_4$ to the exact Bessel solution amounts 
to only about 15\% of the pre-Juno error bar, and the differences in both $J_4$ and $J_6$ are still smaller than 
the uncertainties  of the current Juno data.  The influence of the error in $J_4$ due to application of \ToFfour on 
the predicted envelope metallicity and core mass of Jupiter can be considered negligible. In the following section, 
I investigate whether this is also the case for $J_2$, the error bar of which is 10$\times$ the current Juno estimate
(Fig.~\ref{fig:polyJ2}).

\section{Interior models and $J_2$}\label{sec:modelsJ2}

\begin{figure}[htb]
\centering
\includegraphics[angle=0,width=0.45\textwidth]{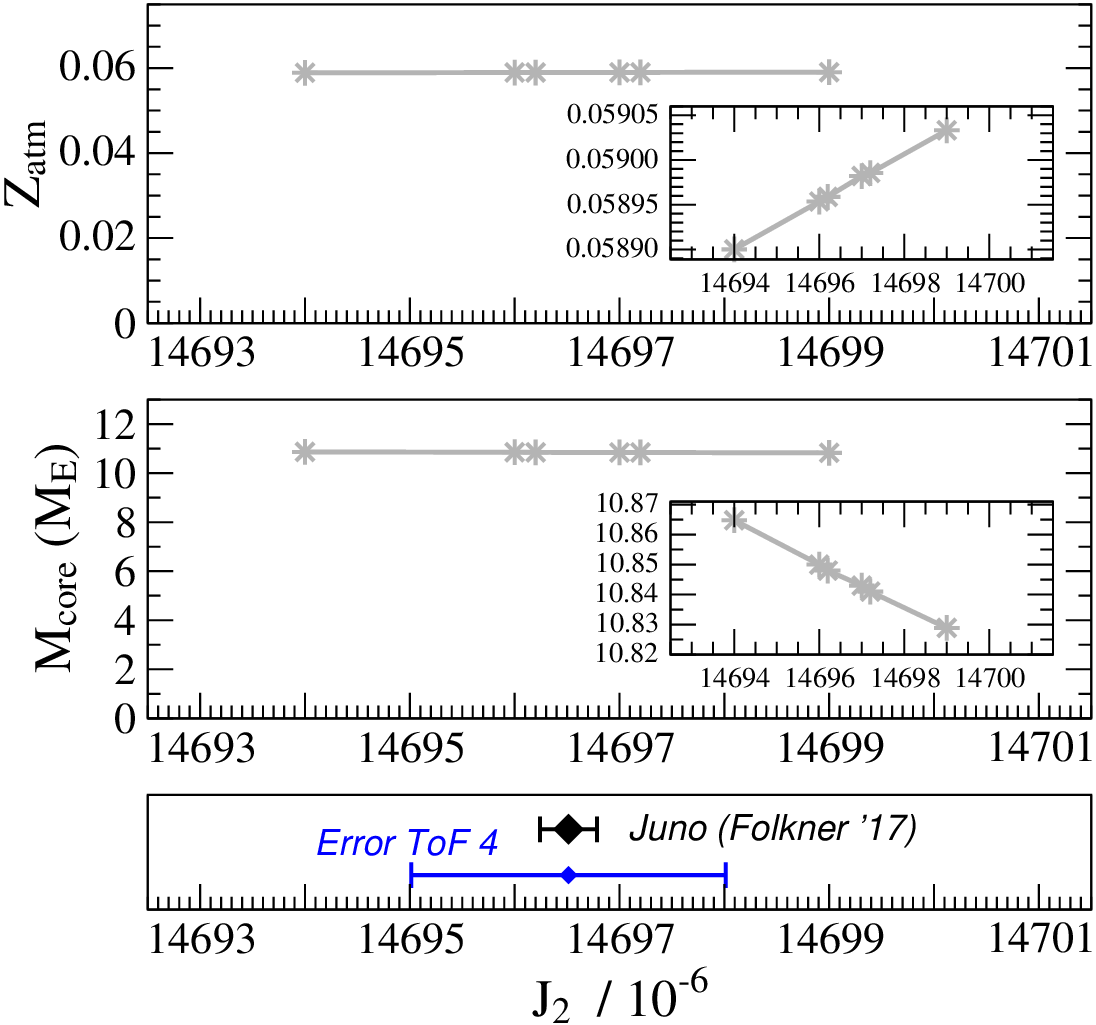}
\caption{Uncertainty in the derived values for core mass and atmospheric metallicity due to the assumed uncertainty 
in $J_2$ for interior models assuming constant envelope-$Z$. For $\Delta J_2$ about $10\times$ the current Juno uncertainty, 
corresponding to $2\times$ the estimated error from  applying \ToFfour (\emph{bottom panel}) the core mass 
uncertainty amounts to $\sim 0.04\ME$ (\emph{middle panel}), while the uncertainty in $Z_{\rm atm}$ is found 
to be less than 0.0002 (\emph{upper panel}). These uncertainties are small.}
\label{fig:J2_McZ}
\end{figure}

The observed value of $J_2$ allows for insight to the internal structure of Jupiter as different internal density 
distributions may yield different values of $J_2$ to be compared against the observed one. 
In this Section I investigate how sensitive that dependence is. In particular, we consider the resulting 
uncertainty in the derived core mass ($\Mcore$) and atmospheric metallicity ($\Zatm$) due to the technical uncertainty 
in $J_2$ which results from applying \ToFfour to compute the gravitational harmonics. For this purpose, simple models 
are computed for which I assume a constant metallicity throughout Jupiter's envelope.  Although further details of 
the procedure do not influence the resulting quantities we are interested in (the uncertainties), I give them for 
completeness: the  envelope is separated into an outer, He-poor part of helium abundance $Y_{\rm atm}=0.238$ in agreement 
with the Galileo entry probe value, and a He-rich inner envelope that accounts for the remaining helium to yield 
a total He/H mass ratio of 0.275 in agreement with estimates for the protosolar cloud. The transition takes place 
at  pressure $\Ptrans=8\:$Mbar. The envelope adiabat runs through the temperature-pressure point of 423 K at 22 bars 
as measured by the Galileo entry probe. At the outer boundary at 1 bar this yields $T_1=170$ K, which I adopt as 
the outer boundary condition for the Jupiter models. Figure \ref{fig:J2_McZ} shows the resulting uncertainties 
in $\Mcore$ and $\Zatm$ as a function of the  assumed value of $J_2$.

\begin{figure}[hbt]
\centering
\includegraphics[angle=0,width=0.45\textwidth]{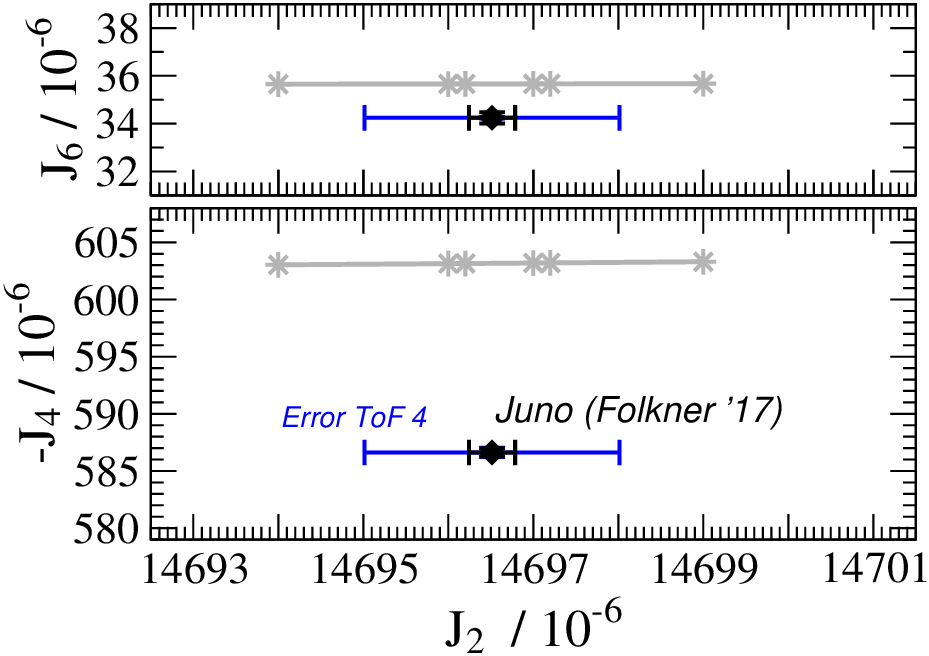}
\caption{Same as Figure \ref{fig:J2_McZ} but for $J_4$ (lower panel) and $J_6$ (upper panel).}
\label{fig:J2_J4J6}
\end{figure}

According to Figure \ref{fig:J2_McZ}, the error in $J_2$ of about $2\times 10^{-4}$ due to applying ToF$\,$4 maps to 
an uncertainty of $0.02\:\ME$ in Jupiter's core mass and 0.0001 in $Z_{\rm atm}$. Furthermore, an uncertainty of
0.1\% in $J_2$ (twice the horizontal length of grey lines) would imply an uncertainty of $\sim 0.1\ME$ in core mass
and $0.0004$ in $\Zatm$.  Thus, the uncertainties due to applying  \ToFfour can be considered tiny compared to 
the accuracy in internal structure properties we are interested in, which is about $10\%$ (e.g., $\sim 1\ME$ 
in core mass). Moreover, the uncertainty from this source of error is clearly smaller than the uncertainty due 
to the material input physics like the EOS, which is at best of the order of 1\%.  
Furthermore, the error due to applying \ToFfour amounts to only 0.04\% in $J_4$ and 0.03\% in $J_6$ 
(Figure \ref{fig:J2_J4J6}). From Figures \ref{fig:J2_McZ} and \ref{fig:J2_J4J6} I therefore conclude that ToF$\,$4 yields quantitatively 
useful density distributions for Jupiter.

\section{Results for Jupiter}\label{sec:models3L}

In this Section I construct models that aim to match the tight current Juno constraints on $J_2$ and $J_4$ and 
are based on H/He-REOS.3 (\citealp{Becker14}, hereafter B14). In Section \ref{sec:McZZ} I assume rocky cores and rigid 
rotation,  while in Section \ref{sec:McZZ_xRz} I assume diluted cores or take into account the shift due to winds. 

\subsection{Models with solid cores and rigid rotation}\label{sec:McZZ}

The models in this Section are three-layer models and constructed as in \citet{Nettelmann12}, hereafter N12. 
The only but important difference to the models of Section \ref{sec:modelsJ2} is that three-layer models allow 
for different heavy element abundances in the two envelopes, so that two free parameters ($Z_1=\Zatm$ 
in the outer and $Z_2$ in the inner envelope) are available for adjusting the two low-order harmonics $J_2$ and $J_4$. 
If this can be achieved and if in addition $Z_1\geq 2\times$ solar, consistent with the observed heavy noble gas 
abundances in Jupiter's atmosphere, I consider a model as acceptable for Jupiter. 

\begin{figure}[hbt]
\centering
\includegraphics[angle=0,width=0.45\textwidth]{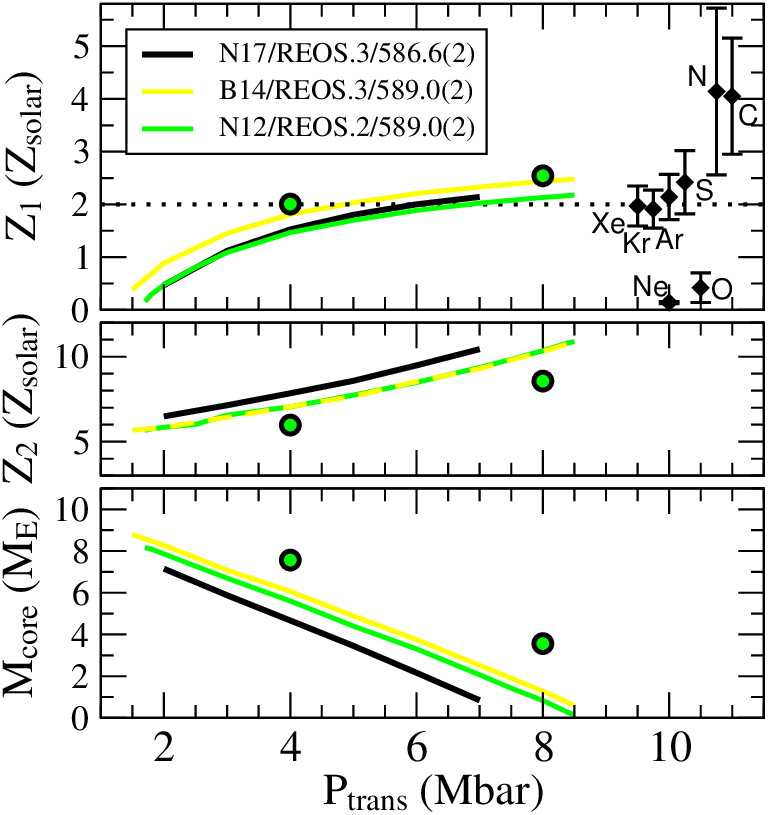}
\caption{ToF$\,$4 based three-layer Jupiter models that are designed to match observed 
$J_2$ and $J_4$ values. This work: \emph{black solid}, B14: \emph{yellow},
N12: \emph{green}. Measured atmospheric particle abundances of the elements as labeled 
are scaled by their protosolar particle abundance values \citep{Lodders03}. I use $Z_{\rm solar}=0.015$.}
\label{fig:McZZ}
\end{figure}

Figure \ref{fig:McZZ} shows these new models as a function of $\Ptrans$. They are similar to the 
\ToFfour based models of N12, who applied H-REOS.2 and He-REOS.1 and of B14, who applied H-REOS.3 and 
He-REOS.3 as in this work, the biggest differences being the narrower range in possible transition pressure and the 
lower $Z_1$ values compared to B14. The latter is mainly a direct consequence of reducing $|J_4/10^{-6}|$ 
from $589$ to $586.6$. 
The lower $Z_1$ values also tend to reduce $J_2$, requiring more heavy elements in the deep interior to compensate 
for that. Sightly higher $Z_2$ values then leave less mass to build the core, so that finally a smaller set of models 
(a smaller range of $\Ptrans$ values for which $\Mcore\geq 0$) is found. In contrast,  
the difference between these models and the N12 results was mainly due to differences in the helium EOS at outer 
envelope pressures. These new results confirm that ab initio H/He-EOSs yield rather low atmospheric metallicities 
for Jupiter. Compared to $\lesssim 1\times$ solar \citep{HM16,WahlHM17}, $\lesssim 2.5\times$ solar (N12), 
$\lesssim 3\times$ solar (B14), I here obtain $Z_{\rm atm} \lesssim 2\times$ solar, out of which 
acceptable models have $\Ptrans=6$--7 Mbar.

An  inaccuracy in $J_4$ of about 2.4/600 (0.4\%), compare black and yellow curves in Fig.~\ref{fig:McZZ}, 
seems to induce a rather large uncertainty of $\Delta\Mcore=2\:\ME$ in core mass; but a 0.1\% uncertainty in $J_4$ might 
still lead to $\Delta\Mcore=0.5\:\ME$ for three-layer models. However, this estimate is probably a far upper bound as the 
models in B14 weere computed with a smaller number of grid points of $N\sim 2000$
compared to $\sim 12,000$ in this work.

\begin{figure}[hbt]
\centering
\includegraphics[angle=0,width=0.46\textwidth]{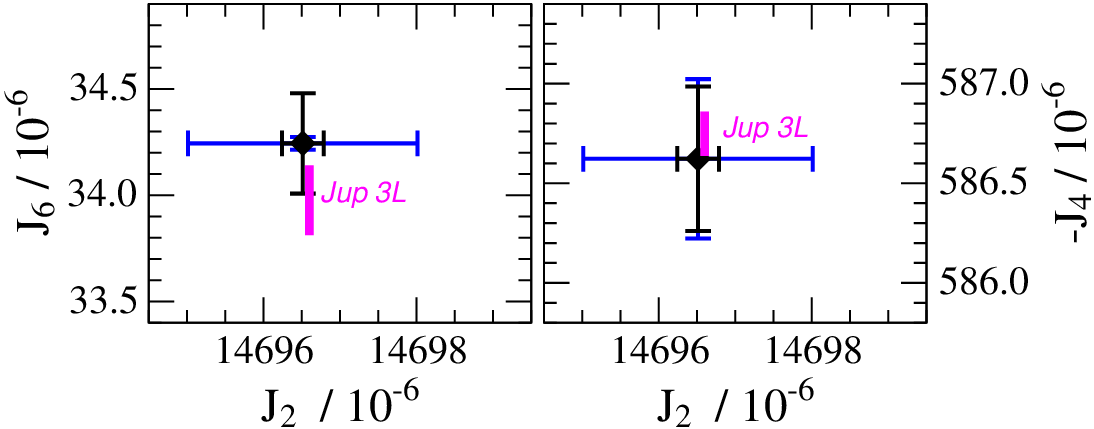}
\caption{(\emph{Left panel:}) Resulting ToF$\,4$ based $J_6$ values of Jupiter models (\emph{magenta}). 
The $J_4$ value of these models (\emph{magenta, right panel}) was adjusted to match the Juno $J_4$ measurement. 
\emph{Black diamonds}: Juno measurements, \emph{blue error bars}: estimated uncertainty due to applying \ToFfour 
according to Figures \ref{fig:polyJ2}--\ref{fig:polyJ6}.} 
\label{fig:Jup3L_J4J6}
\end{figure}

Figure \ref{fig:Jup3L_J4J6} compares the resulting $J_6$ value of the ToF$\,$4 based Jupiter models 
that match the Juno $J_2$, $J_4$ values to the Juno measurement of $J_6$. Models with $P_{\rm trans} = 4$--7 
Mbar are within the observational uncertainty of $J_6$, while models with lower transition pressures are 
within $2\sigma$ of the observational uncertainty. The computational error from \ToFfour is much 
smaller than that.

\subsection{Models with diluted cores or zonal winds}\label{sec:McZZ_xRz}

Ab initio H/He EOS based Jupiter models with rock-ice cores and without zonal winds become notoriously low in 
atmospheric heavy element abundances. On the other hand, diluted cores
have been found to enhance $\Zatm$ by up to 50\% \citep{ForNett10}, while zonal winds direcly affect the values 
of $J_2$ and $J_4$ to be matched by rigidly rotating models \citep{Militzer08,CaoStev17}. While precise predictions 
on the dynamic contributions $\Delta J_{2n}$ to the observed values depend on the differential rotation pattern 
and their mathematical description \citep{Kaspi10,Zhang15,CaoStev17}, it is predicted that the effect on the 
low-order $J_{2n}$ increases with the depth of the winds \citep{Kaspi10,CaoStev17}, that the effect on the 
low-order $J_{2n}$ is small and in the direction of reducing their absolute values 
\citep{Hubbard99,Kaspi10,CaoStev17}.  Here I calculate Jupiter models as in Section \ref{sec:McZZ} but by assuming 
a diluted core of rock mass fraction $Z_{3,\rm\,Rocks}=0.2$, the rest being inner mantle 
material, and by including zonal wind corrections as proposed by \citet{CaoStev17} for half-amplitude widths 
(HAWD) values of 0.8 and 0.9. The latter quantity is defined as the distance to the rotation axis where the 
azimuthal wind velocity has weakened by a factor of two from its maximum value farther out. 

\begin{figure}[hbt]
\centering
\includegraphics[angle=0,width=0.45\textwidth]{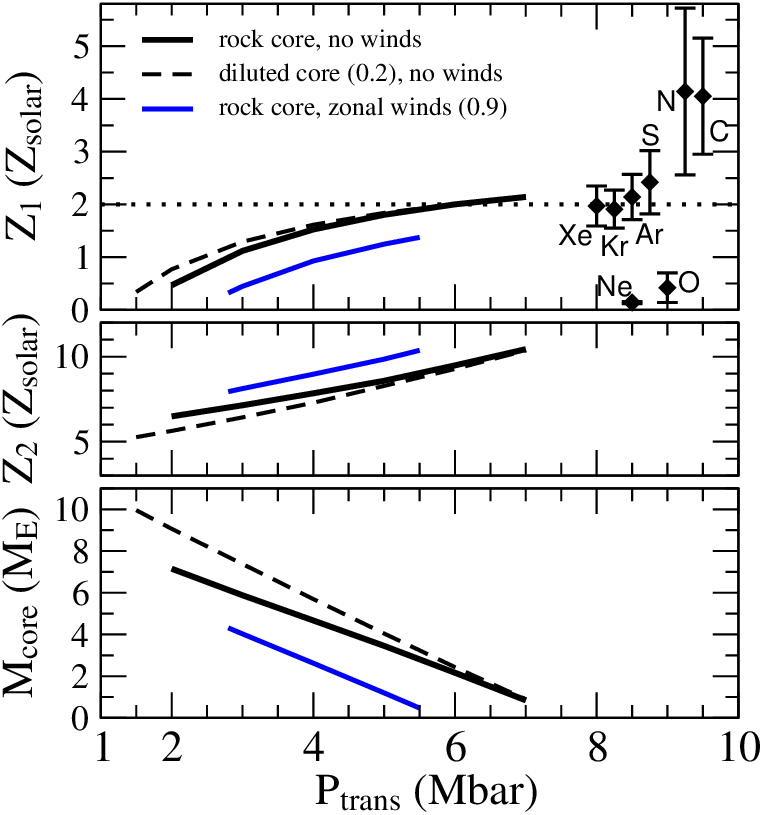}
\caption{Jupiter models with diluted cores of central rock mass fraction $Z_{3,\rm\,Rocks}=0.2$ (black dashed) or with 
zonal winds according for HAWD=0.9 \emph{(solid blue)}. The \emph{solid black} curves are the same as in Figure
\ref{fig:McZZ}. I use $Z_{\rm solar}=0.015$.}
\label{fig:McZZ_xRz}
\end{figure}

As shown in Figure \ref{fig:McZZ_xRz}, the zonal wind corrections lead to lower values in $Z_1$ and $\Mcore$.
This is not a surprise, since the absolute values of $J_2$ and $J_4$ are reduced and thus demand a smaller mass
density in the planet where they are most sensitive, which is near $P\sim 1$~Mbar in the outer envelope.
This behavior is in line with the observation of \citet{Militzer08} who, in order to \emph{enhance} the resulting 
envelope $Z$ value,  suggest zonal wind effects on $J_2$ and $J_4$ in the \emph{opposite} direction of what zonal 
wind models that fit the observed wind speeds predict.
 
For HAWD=0.8 I do not obtain any acceptable Jupiter model; the $\Delta J_{2n}$ are too large: both  $Z_1$ and $M_{core}$ would 
become negative. For for HAWD=0.9, there is a restricted range of solutions at $\Ptrans=3$--5.5 Mbar, 
for which $Z_1$ barely reaches $1.5\times$ solar. Adiabatic H/He-REOS.3 based Jupiter models thus suggest 
the vertical extend of the winds to be less than 0.9 $\RJ$ ($\sim 7000$ km).

Assuming a diluted core and adiabatic envelopes, the $Z_1$ value can be lifted, but only to less than its maximum 
value obtained for core-less models. The enhancement in $Z_1$ can indeed reach up to 50\% for the largest core mass found
here, but then the base $Z_1$ value is small anyway. Therefore, as Figure \ref{fig:McZZ_xRz} shows, diluted cores 
do not significantly enhance $Z_{atm}$ for H/He-REOS based models, but are helpful 
for larger core models \citep{WahlHM17} such as obtained with the DFT-MD EOS of \citet{MH13}.

\section{High-order gravitational harmonics}\label{sec:highJs}

To compute the high-order gravitational harmonics of models that match the observed Juno values for $J_2$ and
$J_4$ I use the density distributions of the models from Section \ref{sec:models3L} and apply the CMS method to 
them 
\footnote
{One could of course use CMS method right from the start; however, my current implementation of the CMS method runs 
orders of magnitudes slower than my implementation of the ToF method. Given the number of iterations necessary to 
fit both $J_2$ and $J_4$ according to the procedure outlined in \citet{Nettelmann11}, I evaluated the accuracy 
gained not worth the enormous computational extra effort.
}.

\begin{figure}[hbt]
\centering
\includegraphics[angle=0,width=0.45\textwidth]{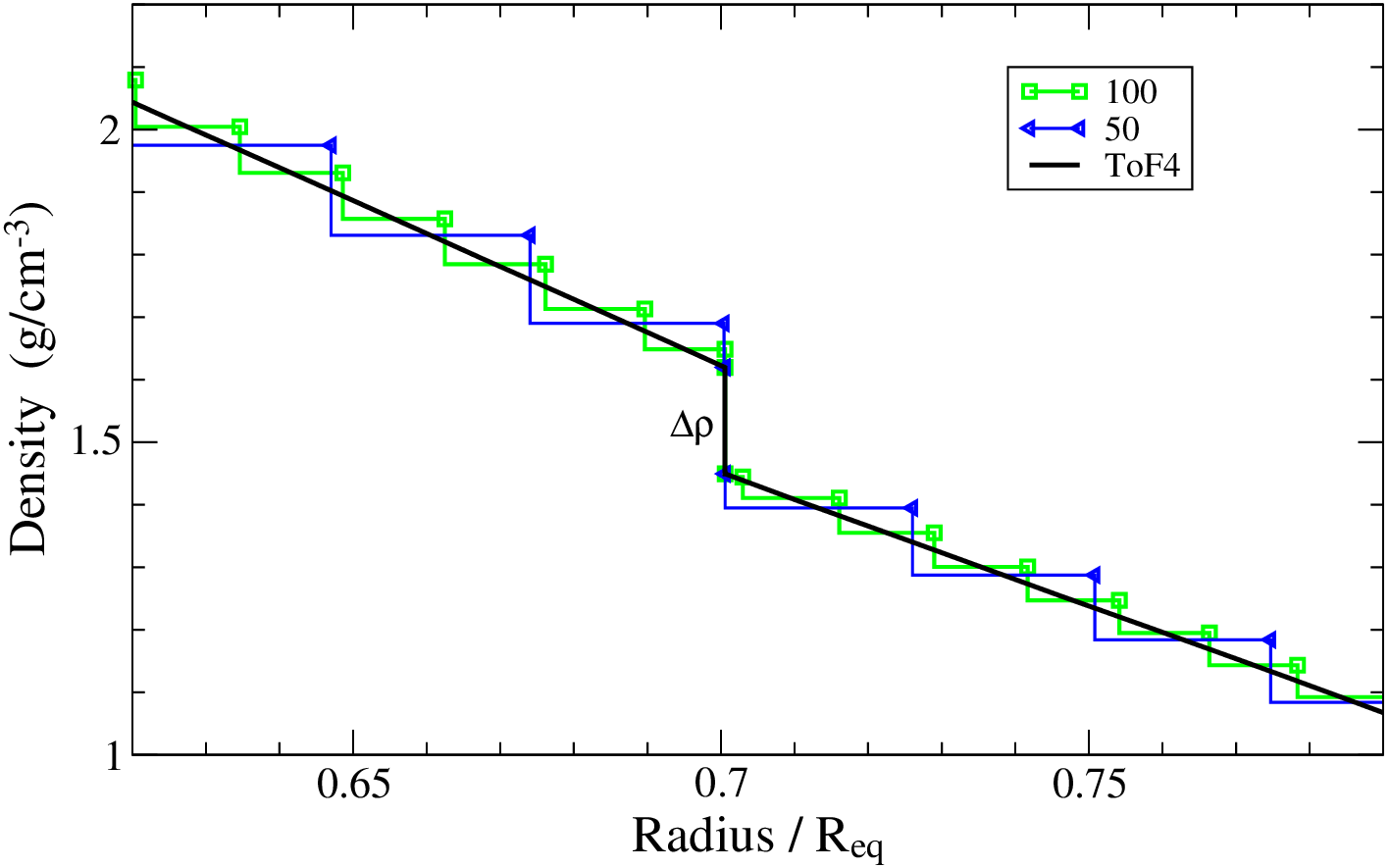}
\caption{Conversion of the \ToFfour based density profile (\emph{black}) to a discrete grid as required by 
CMS method, here illustrated for $N=50$ (\emph{blue}) and $N=100$ (\emph{green}).}
\label{fig:rho_r_convert}
\end{figure}

For that purpose, I convert the ToF-based density profile as a function of mean radius of an equipotential surface 
to a density profile as a function of equatorial radius of same equipotential surface 
 using the ToF$\,$4 based figure functions. Then I reduce the number of 
radial grid points from $N\sim 12,000$ to $N\sim 1000$ by assuming  a radial spacing that decreases 
continuously from the middle to the boundaries 
At layer boundaries, the jump in density is conserved as illustrated in Figure \ref{fig:rho_r_convert}.

\begin{table*}[thb]
\caption{High-order moments of different models for rigidly rotating Jupiter}
\label{tab:highJs}     
\tiny
\begin{center}             
\begin{tabular}{rcccccccc} 
\hline\hline                
$J_{2n}\quad$ 		& Juno 								& ToF-4 	& CMS-1000$^{a,b}$	& CMS-1000$^{a,b}$ & Polytrope$^{c}$ 	& CMS [HM16]& Juno 0.9$^{d}$ & CMS-1000$^{e}$\\    
     					& [F17]  							&J17-3a		&J17-3a/b		&J17-6a/b	&[WH16] 	& DFT-MD 7.13		& [CS17]	& J17-4z \\    \hline                     
$J_2/10^{-6}$ 		& 14696.514 $\pm$ 0.272	& 14696.6	&	$\begin{array}[t]{c}14698.30^{a}\\14696.50^{b}\end{array}$ 	
	& $\begin{array}[t]{c}14698.24^{a}\\14696.64^{b}\end{array}$ 
																														& 13988.15	& 14696.43 	& 14690.68 	& 14692.26\\      
$-J_4/10^{-6}$ 	& 586.623 $\pm$ 0.363 	& 586.64	&	586.65		& $\begin{array}[t]{c}586.62^{a}\\586.63^{b}\end{array}$ 			
																														& 531.83	& 596.05	& 581.91	& 582.00\\
$J_6/10^{-6}$ 		& 34.244 $\pm$ 0.236 	 	& 34.09 	&	34.21		& 34.42  	& 30.12	& 35.15 	& 31.75	& 33.85\\
$-J_8/10^{-6}$ 	& 2.502 $\pm$ 0.311 		& 2.732		&	2.460		& 2.491  	&	2.132	& 2.546 	& 1.335     & 2.433\\
$J_{10}/10^{-7}$ &...									& ...		& 2.021			& 2.057		& 1.741 	& 2.10 	& ...		& 1.999\\     
$-J_{12}/10^{-8}$ &...								& ...		& 1.821 		& 1.860		& 1.568 	& ...		& ...		& 1.801\\
$J_{14}/10^{-9}$ &...									& ...		& 1.755			& 1.797		& 1.518 	& ...		& ...		& 1.736\\
$-J_{16}/10^{-10}$ &...								& ...		& 1.781			& 1.827		& 1.552 	& ...		& ...		& 1.762\\
$J_{18}/10^{-11}$ &...								& ...		& 1.883  		& 1.934		& 1.656 	& ...		& ...		& 1.862\\
\hline                     
\end{tabular}
\end{center}
Refs.: [F17]=\cite{Folkner17}, [WH16]=\cite{WH16}, [HM16]=\cite{HM16}, [CS17]=\cite{CaoStev17}; 
${}^{a}$: $J_2$ and $J_4$ fitted to Juno data using ToF$\,$4; 
${}^{b}$: same as $(a)$ but for $J_2$ value to be fitted shifted by the difference \ToFfour$-$CMS, values 
are displayed only where different from $(a)$; 
${}^{c}$: Bessel solution for rigid rotation, 
${}^{d}$: Juno data corrected for zonal winds based on Bessel solution for the $\Delta J_{2n}$ of polytropic model
with wind depth HAWD$=0.9$.  
${}^{e}$: same as $(a)$ but using the $J_2$ and $J_4$ values from $(d)$.
\end{table*}


Finally, Table \ref{tab:highJs} presents my results for the low- and high-order $J_{2n}$ values of two 
models of Section \ref{sec:McZZ}, i.e.~for $\Ptrans=3$ Mbar (model J17-3a) and 6 Mbar (model J17-6a), 
and for one model which accounts for deep zonal winds through the corrections to $J_2$ and $J_4$ for HAWD=0.9 
from Section \ref{sec:McZZ_xRz} (model J17-4z). 
Resulting moments of order $\geq 6$ are not affected by the above described procedure within the number 
of digits given in Table \ref{tab:highJs}. This is shown by model variant $(b)$, where the $J_2$ value to be 
fitted was shifted by the difference \ToFfour$-$CMS according to model variant $(a)$.  
The results are compared to the exact polytrope solution \citep{WH16}, to the DFT-MD-7.13 Jupiter model 
of \citet{HM16}, to the Juno measurements of \citet{Folkner17}, and to that data but corrected for deep 
zonal winds as proposed by \citet{CaoStev17}.

Perhaps most interestingly, the resulting values for $J_6$ and $J_8$ for rigidly rotating Jupiter are within 
the current Juno observational error bars. This may indicate that the winds are shallow. The $J_8$ value of 
model J17-4z is also within the observational error bar and the reduction of its $|J_8|$ value by few percent 
is much less than the few 10\% estimate of \citet{CaoStev17} whose $\Delta J_8$ estimate peaks for HAWD=0.9 compared to 
deeper (0.8) or shallower (0.975) depths. Furthermore, the high-order $J_{2n}$ values of $n=1$ polytropic 
Jupiter differ by about 10\% from the physical EOS  based Jupiter models. Thus it is
important to provide the latter class of models as done in this work and in \citet{HM16}.

\section{Conclusions}\label{sec:conclusion} 

To infer Jupiter's internal density distribution, the relative accuracy in the computed values of $J_2$ and $J_4$ 
should be of order 0.1\% (Figures \ref{fig:J2_McZ} and \ref{fig:McZZ}). According to $n=1$ polytropic models, 
\ToFfour can provide this accuracy, while \ToFthree only for $J_2$ (Figures \ref{fig:polyJ2} and \ref{fig:polyJ4}). 
The error in $J_2$ ($J_4$, $J_6$) due to applying \ToFfour is about $10\times$ ($1/2\times$, $1/3\times$) 
the current \emph{Juno} estimates for these parameters. I conclude that these uncertainties are nevertheless 
sufficently small for predicting Jupiter's internal density distribution. Other uncertainties, such as the thermal
state, perhaps as a result of helium rain, may induce larger unknowns in our understanding of Jupiter 
\citep{Nettelmann15,HM16,Mankovich16,WahlHM17}.

The computed values of $J_6$ and $J_8$ of rigidly rotating Jupiter suggests that zonal wind are restricted
to regions well above a depth of 7000 km ($0.9\RJ$). Furthermore, application of the more accurate CMS method in combination 
with the physical EOS H/He-REOS.3 is found to yield higher-order $|J_{2n}|$ values that are 10\% higher than the 
prediction from the polytropic model (Table \ref{tab:highJs}). 

Still, the internal structure of Jupiter remains poorly constrained. Further insight might be gained 
from a Juno measurement of the fluid Love number $k_2$ and its consideration in three-dimensional models
for the gravity field \citep{Wahl17saturn}.

\acknowledgements

I thank Naor Movshovitz for sharing results for linear density models with the CMS method,
and Ronald Redmer and Ravit Helled for inspiring discussions. This work was supported by the DFG grant
NE1734/1-1 of the German Science Foundation.

\bibliography{refsJup}

\appendix

\section{Linear Density models with CMS}\label{sec:linear}

\citet{Hubbard13} provides the gravitational harmonics for a linear density model with Jupiter-like parameters
and $N=128$. I use that model to test my implementation of the CMS method. The agreement is excellent for the 
given number of digits (7 in H13), see Table \ref{tab:linear}. However, a model with only $N=128$ is not yet 
converged as the same linear density model with $N=512$ shows. Therefore, in order to obtain this good agreement 
with the linear density model of H13, I had to chose the spacings in equatorial radius $d\lambda_i$ and the 
dimensionless density jumps $\delta_i$ exactly as in H13. In particular (N.~Movshovitz, pers.~comm.), 
$\lambda_0=1$, $\lambda_{N}=0$, $\delta_0=\delta_{N}=0$, $d\lambda= 1/(N-1)$, 
$\lambda_{1} = \lambda_{0} - d\lambda/2$,  
$\lambda_{i} = \lambda_{i-1} - d\lambda$ for $i>1$, 
and $\delta_{i} = d\lambda_{i}$ for $i\geq 1$. 
All other parameters were chosen as in H13, in particular $q=0.088822426$, $R_{eq}=71492\,$km, 
$\rm GM=126686536\:km^3/s^2$.

\begin{table}[hhh]
\caption{CMS results for linear density model}
\label{tab:linear}     
\centering             
\begin{tabular}{rccc} 
\hline\hline                
$J_{2n}$ & CMS-128 & CMS-128 & CMS-512  \\    
         & [H13] & \multicolumn{2}{c}{this work}\\
\hline                     
$J_2/10^{-2}$ 		& 1.4798138 & 1.47981376 & 1.47978941  \\      
$-J_4/10^{-4}$ 	& 5.9269129 & 5.92691294 & 5.92726570  \\
$J_6/10^{-5}$ 		& 3.4935680 & 3.49356798 & 3.49433822  \\
$-J_8/10^{-6}$ 	& 2.5493209 & 2.54932089 & 2.55049835  \\
$J_{10}/10^{-7}$ & 2.1308951 & 2.13089515 & 2.13255938  \\
$-J_{12}/10^{-8}$ & 1.9564143 & 1.95641425 & 1.95871536  \\
$J_{14}/10^{-9}$ & 1.9237724 & 1.92377252 & 1.92693981  \\
\hline                     
\end{tabular}\\
All parameters are chosen as in [H13]=\citet{Hubbard13}.
\end{table}




\section{ToF to 4th order coefficients}

I summarize the Theory of Figures of \citet{ZT78} and then give the coefficients up to 4th order. Consider a spheroidal planet in hydrostatic equilibrium of density distribution $\rho(r,\vartheta)$ which is symmetric with respect to the axis of rotation and the equatorial plane.  As a result, there is no dependence on azimuthal angle $\varphi$, and only even indices in the spherical harmonics expansions survive.
In this two-dimensional problem, a surface of constant total potential $U$ only depends on polar angle
$\vartheta$. Different such  surfaces $r_l(\vartheta)$ are labeled by the level parameter $l$. 
In ToF method according to \citet{ZT78}, $l$ is taken to be the mean radius of the respective equipotential surface as defined by the condition of equal volume, 
$(4\pi/3)\:l^3=2\pi\int_{-1}^1 d\cos\vartheta\int_0^{r_l(\vartheta)} dr'\:r'^2$. 
Furthermore, any dependence on $(r,\vartheta)$ is replaced by dependence on $(l, \vartheta)$ through the expansion of $r_l(\vartheta)$ into a series of Legendre polynomials $P_n(\cos\vartheta)$ according to
\begin{equation}\label{eq:rltheta}
	r_l(\vartheta)=l\left(1+\sum_{n=0}^{\infty} s_{2n}(l)\:P_{2n}(\cos\vartheta)\right)\quad,
\end{equation} 
where the $s_{2n}(l)$ are the \emph{figure functions}.  The first-order deviation from a spherical shape is described by $s_2$, while $s_0$ can be determined with the help of the equal-volume condition to
\begin{equation}\label{eq:s0}
s_0 = -\frac{1}{5}\,s_2^2 - \frac{2}{105}\,s_2^3 - \frac{1}{9}\,s_4^2 - \frac{2}{35}\,s_2^2 s_4\quad.
\end{equation}
The figure functions $s_{2n}$ are of $n$-th order except $s_0$ which is of 4th order. In the following, I abbreviate 
the expression in parenthesis in  Eq.~(\ref{eq:rltheta}) by $(1+\Sigma)$ and set $\mu=\cos\vartheta$.

The total potential is composed of the gravitational potential  $V(\vec{r}) = -G\int d^3r' \rho/|\vec{r}'-\vec{r}|$
while the centrifugal potential reads $Q=-\frac{1}{2}\omega^2 r^2 \sin^2\vartheta$. In ToF it is convenient to capture the centrifugal term due to the planetary rotation of angular rotation rate $\omega$ by the small parameter 
$m=\omega^2 R_{\rm m}^3/GM$, where $R_{\rm m}$ is the mean radius of the outermost level surface.
After expanding $V$ and $Q$ into series of Legendre polynomials and replacing $r$ by Eq.~(\ref{eq:rltheta}) one can write 
\begin{equation}
	U(l,\vartheta) = - \frac{4\pi}{3}\,G\,\bar{\rho}\,l^2\sum_{k=0}^{\infty} A_{2k}(l)\,P_{2k}(\mu)
	\quad,
\end{equation}
where $\bar{\rho}$ denotes the mean density $3M/(4\pi R_{\rm m}^3)$. On equipotential surfaces, $dU/d\theta=0$ and thus $A_{2k}\equiv 0$ for $k>0$. This property is used to determine the $s_{2n}$, while $A_0$ yields the total potential. One finds
\begin{eqnarray}
	A_0 &=& \left(1 + \frac{2}{5}\,s_2^2 -\frac{4}{105}\,s_2^3 + \frac{2}{9}\,s_4^2 + \frac{43}{175}\,s_2^4 
		- \frac{4}{35}\,s_2^2 s_4\right) S_0 
\nonumber\\
	&& \hspace{-0.8cm} + \left(-\frac{3}{5}\,s_2 +\frac{12}{35}\,s_2^2 -\frac{234}{175}\,s_2^3 
		+ \frac{24}{35}\,s_2 s_4\right)\,S_2 
	 + \left(-\frac{5}{9}\,s_4 + \frac{6}{7}\,s_2^2\right) S_4 
\nonumber\\
	&& \hspace{-0.8cm} + \:S_0'\: + \left(\frac{2}{5}\,s_2 + \frac{2}{35}\,s_2^2 + \frac{4}{35}\,s_2 s_4 
		- \frac{2}{25}\, s_2^3\right)\,S_2' + \left( \frac{4}{9}\,s_4 + \frac{12}{35}\,s_2^2\right) S_4'
\nonumber\\
	&& \hspace{-0.8cm} + ~ \frac{m}{3}\left(1 - \frac{2}{5} s_2 - \frac{9}{35} s_2^2 - \frac{4}{35} s_2 s_4  
		+ \frac{22}{525} s_2^3\right) \quad.
	\label{eq:A0}
\end{eqnarray}
The functions $S_n$ and $S_n'$ will be defined below.

\subsection{From $V$ to $D_n,D_n'$ and further to $S_n$, $S_n'$, and $J_n$}

The gravitational potential at a location $(r,\vartheta)$ in the planet separates into an external potential  
$r>r'$ due to the mass distribution $\rho(r',\vartheta')$ interior to $r$ and an internal potential $r<r'$ due to the mass distribution $\rho(r',\vartheta')$ exterior to $r$. The multipole expansion of $V$ reads 
\begin{equation}\label{eq:graviV}
	V(r,\vartheta) = -\frac{G}{r}\sum_{n=0}^{\infty}\left(r^{-2n}D_{2n}(r) + r^{2n+1}D_{2n}'(r)\right)P_{2n}(\mu)\:.
\end{equation}
Using Eq.~(\ref{eq:rltheta}), the volume integrals $D_n$ of the external and $D_n'$ of the internal gravity field expansion take the form \footnote{In the representation by equipotential surfaces, $r<l'$ can happen for the external field and $r>l'$ for the internal field. This does not pose a problem here \citep{ZT78} as long as the $m$-value is sufficiently small \citep{Hubbard14}.} 
\begin{eqnarray}
	D_n(l) &=& \frac{2\pi}{n+3}\int_0^l\!dl'\, \rho(l')
	\int_{-1}^1 d\mu'\, P_n(\mu')\:\frac{d\,r^{n+3}}{dl}
	\nonumber\\
	D'_n(l) &=& \frac{2\pi}{2-n}\int_l^{R_{\rm m}}\!dl'\: \rho(l')
	\int_{-1}^1 d\mu'\,P_n(\mu')\: \frac{d\,r^{(2-n)}}{dl'} \quad(n\not=2)
	\label{eq:Dnp_l}
	\nonumber\\
	D'_2(l) &=& 2\pi\int_l^{R_{\rm m}} dl' \: \rho(l')
	\int_{-1}^1\!d\mu'\: P_2(\mu')\:\frac{d\,\ln r}{dl'} \quad.
	\label{eq:DDp}
\end{eqnarray}
With $z:=l/R_{\rm m}$, their dimensionless form is defined as
\begin{equation}	\label{eq:Sn_Dn}
  	S_n(z) = \frac{3}{4\pi\bar{\rho}\,l^{n+3}}\,D_n(l)\quad,\quad
	S'_n(z) = \frac{3}{4\pi\bar{\rho}\,l^{2-n}}D'_n(l)\quad,
\end{equation}
and can be written as
\begin{eqnarray}
	S_n(z) &=& \frac{1}{z^{n+3}} \int_0^{z}\! dz'\: \frac{\rho(z')}{\bar{\rho}}\,\frac{d}{dz'}[z'^{\,n+3}f_n(z')]
\nonumber\\
	S'_n(z) &=& \frac{1}{z^{(2-n)}} \int_{z}^1\! dz'\: \frac{\rho(z')}{\bar{\rho}}\,\frac{d}{dz'}[z'^{\,2-n}f'_n(z')]
\nonumber \\
	S_0(z) &=& \frac{m(z)}{M z^3}\quad.
	\label{eq:Sn_def} 
\end{eqnarray} 
After application of partial integration and assuming $d\rho/dz$ to be finite, the $S_{n}$, $S_n'$ adopt the convenient form for numerical evaluation
\begin{eqnarray}\label{eq:SnSnp_num}
S_n(z) &=& \frac{\rho(z)}{\bar{\rho}}\,f_n(z) - \frac{1}{z^{n+3}}\int_0^z \frac{d\rho}{\bar{\rho}}\,z'^{\,n+3}f_n(z')
\\
S_n'(z) &=& -\frac{\rho(z)}{\bar\rho}\,f_n'(z) + \frac{1}{z^{2-n}}\left(\frac{\rho(1)}{\bar{\rho}}\,f_n'(1)
		- \int_z^1 \frac{d\rho}{\bar{\rho}}\,z'^{\,2-n}f'_{n}(z')\right)
\:.\nonumber
\end{eqnarray}
with 
\begin{eqnarray}
f_n(z) &=& \frac{3}{2(n+3)}\int_{-1}^1\! d\mu\: P_n(\mu)\:(1+\Sigma)^{n+3}\quad,
\nonumber\\
f'_n(z) &=& \frac{3}{2(2-n)}\int_{-1}^1\! d\mu\: P_n(\mu)\:(1+\Sigma)^{2-n} \quad (n\not=2)\,,
\nonumber\\
f'_2(z) &=& \frac{3}{2}\int_{-1}^1\!d\mu\: P_n(\mu)\:\ln(1+\Sigma)\quad.
\label{eq:fnfnp}
\end{eqnarray}
By expressing powers of $(1+\Sigma)$ in terms of the binomial series expansions, and by further expanding powers of $\Sigma$ into linear series of Legendre polynomials, and by making use of $\int_{-1}^1 d\mu\: P_n(\mu)\,P_m(\mu)=0$ for $n\not=m$, 
the integrals in Eqs.~(\ref{eq:fnfnp}) can be solved analytically. The results for $f_n(z)$ and $f_n'(z)$ are provided in  Eqs.~(\ref{eq:fn}) and (\ref{eq:fnp}). Accordingly, the integrals $S_{2n}$ and $S_{2n}'$ are of $n$-th order.
Finally, the gravitational harmonics are obtained as 
\begin{equation}\label{eq:J2n}
	J_{2n}=-(R_{\rm m}/R_{\rm eq})^{2n}\,S_{2n}(1)\:.
\end{equation}

\subsection{Coefficients in $A_{2n}$ for computing the $s_{2n}$}

Below I give the coefficients that are of 4th order or lower \emph{after} multiplication with $m$, $S_{2n}$, 
or $S_{2n}'$ as occurring in the respective equations. They were generated by a C++ program written by myself in 2004. 

\begin{eqnarray}
	A_2 &=& \left(-s_2 + \frac{2}{7}\,s_2^2  +\frac{4}{7}\,s_2 s_4 - \frac{29}{35}\,s_2^3 
		+ \frac{100}{693}\,s_4^2 + \frac{454}{1155}\,s_2^4 \right.
 	\nonumber\\
	&& - ~ \left. \frac{36}{77}\,s_2^2 s_4 \right) S_0  + \left(1-\frac{6}{7}\,s_2  - \frac{6}{7}\,s_4 
		+ \frac{111}{35}\,s_2^2 - \frac{1242}{385}\,s_2^3 \right.
	\nonumber \\
	&& + ~ \left. \frac{144}{77}\,s_2 s_4\right)S_2 + \left( -\frac{10}{7}\,s_2 - \frac{500}{693}\,s_4 
		+ \frac{180}{77}\,s_2^2 \right) S_4
	\nonumber\\
	&& \hspace{-0cm} 	+ \left(1 + \frac{4}{7}\,s_2 +\frac{1}{35}\,s_2^2 + \frac{4}{7}\,s_4 - \frac{16}{105}\,s_2^3 
		+ \frac{24}{77}\,s_2 s_4 \right) S_2'
	\nonumber\\
	&&  \hspace{-0cm} 	+ \left( \frac{8}{7}\,s_2 + \frac{72}{77}\,s_2^2 + \frac{400}{693}\,s_4 \right) S_4'
	\nonumber\\
	&&  \hspace{-0.5cm} + ~ \frac{m}{3}\left(- 1 + \frac{10}{7} s_2 + \frac{9}{35} s_2^2 - \frac{4}{7} s_4 
		+ \frac{20}{77} s_2 s_4 - \frac{26}{105} s_2^3 \right)
	\label{eq:A2}\\
	\parbox{0cm}{\mbox{To lowest order, $s_2\approx -m/3$, thus $s_2$ is of first order in $m$.}} 
	\nonumber
\end{eqnarray}
\begin{eqnarray}
	A_4 &=& \left(-s_4 + \frac{18}{35}\,s_2^2 - \frac{108}{385}\,s_2^3 + \frac{40}{77}\,s_2 s_4 
			+ \frac{90}{143}\,s_2 s_6 + \frac{162}{1001}\,s_4^2 \right.
	\nonumber\\
	&& \left. + ~ \frac{16902}{25025}\,s_2^4 - \frac{7369}{5005}\,s_2^2 s_4\right)S_0
			+ \left( - \frac{54}{35}\,s_2 - \frac{60}{77}\,s_4 + \frac{648}{385}\,s_2^2 \right. 
	\nonumber\\
	&&	 \left. - ~ \frac{135}{143}\,s_6 + \frac{21468}{5005}\,s_2 s_4 - \frac{122688}{25025}\,s_2^3 \right)S_2
	+ \left( 1 - \frac{100}{77}\,s_2 \right.
	\nonumber\\
	&&  - ~ \left. \frac{810}{1001}\,s_4 + \frac{6368}{1001}\,s_2^2 \right) S_4  
	-  \frac{315}{143}\,s_2 \, S_6 + \left( \frac{36}{35}\,s_2 \right.
	\nonumber\\
	&& \hspace{-0cm}  + ~ \left. \frac{108}{385}\,s_2^2 + \frac{40}{77}\,s_4 
	+ \frac{3578}{5005}\,s_2 s_4 - \frac{36}{175}\,s_2^3 + \frac{90}{143}\,s_6\right) S_2'
	\nonumber\\
	&& \hspace{-0cm} + \left( 1 + \frac{80}{77}\,s_2 + \frac{1346}{1001}\,s_2^2 + \frac{648}{1001}\,s_4 \right) S_4'
	+  \frac{270}{143}\,s_2\,S_6'
	\nonumber\\
	&& \hspace{-0.5cm} + ~ \frac{m}{3}\left(-\frac{36}{35} s_2 + \frac{114}{77} s_4  + \frac{18}{77} s_2^2  
		- \frac{978}{5005} s_2 s_4  + \frac{36}{175} s_2^3  - \frac{90}{143} s_6 \right)
	\nonumber\\
	\label{eq:A4}\\
\parbox{0cm}{\mbox{To lowest order, $s_4\sim m\times s_2$, thus $s_4$ is of 2nd order in $m$.}} 
\nonumber
\end{eqnarray}
\begin{eqnarray}
	A_6 &=& \left( -s_6 + \frac{10}{11}\,s_2 s_4 - \frac{18}{77}\,s_2^3 + \frac{28}{55}\,s_2 s_6 + \frac{72}{385}\, s_2^4 
		+ \frac{20}{99}\,s_4^2 \right.
	\nonumber\\
	&& - ~\left. \frac{54}{77}\,s_2^2 s_4 \right) S_0  + \left( - \frac{15}{11}\,s_4 
	+ \frac{108}{77}\,s_2^2 	- \frac{42}{55}\,s_6 - \frac{144}{77}\,s_2^3 \right.
	\nonumber\\
	&& + ~ \left. \frac{216}{77}\,s_2 s_4 \right) S_2 
	+ \left( - \frac{25}{11}\,s_2 - \frac{100}{99}\,s_4 + \frac{270}{77}\,s_2^2\right) S_4 
	\nonumber\\
	&& \hspace{-0cm} + \left(1 - \frac{98}{55}\,s_2 \right) S_6 + \left( \frac{10}{11}\,s_4 + \frac{18}{77}\,s_2^2 
	+ \frac{36}{77}\,s_2 s_4 + \frac{28}{55}\,s_6\right) S_2' 	
	\nonumber\\
	&& + \left( \frac{20}{11}\,s_2 + \frac{108}{77}\,s_2^2 + \frac{80}{99}\,s_4\right) S_4' 	
	+ \left( 1+\frac{84}{55}\,s_2\right) \, S_6' 
	\nonumber\\
	&& \hspace{-0cm} + ~ \frac{m}{3}\left(- \frac{10}{11} s_4 - \frac{18}{77} s_2^2 + \frac{34}{77} s_2 s_4 
	+ \frac{82}{55} s_6  \right)
	\nonumber\\
	\label{eq:A6}\\
	\parbox{0cm}{\mbox{To lowest order, $s_6\sim m\times s_4$, thus $s_6$ is of 3rd order in $m$.}}
	\nonumber
\end{eqnarray}
\begin{eqnarray}
	A_8 &=&  \left( -s_8 + \frac{56}{65}\,s_2 s_6 + \frac{72}{715}\,s_2^4 + \frac{490}{1287}\,s_4^2 
		- \frac{84}{143}\,s_2^2 s_4\right) S_0 
	\nonumber\\
	&& \hspace{-0cm} 	+ \left( - \frac{84}{65}\,s_6 - \frac{144}{143}\,s_2^3 + \frac{336}{143}\,s_2 s_4 \right) S_2
		+ \left( - \frac{2450}{1287}\,s_4 \right.
	\nonumber\\
	&& + ~ \left. \frac{420}{143}\,s_2^2 \right) S_4 - \frac{196}{65}\,s_2\,S_6 + S_8
		+ \left( \frac{56}{65}\,s_6 + \frac{56}{143}\,s_2 s_4\right) S_2' 
	\nonumber\\
	&& + \left( \frac{1960}{1287}\,s_4 + \frac{168}{143}\,s_2^2\right) S_4' + \frac{168}{65}\,s_2\,S_6' + S_8'
	\nonumber\\
	&& + ~ \frac{m}{3}\left(- \frac{56}{65} s_6  - \frac{56}{143} s_2 s_4 \right)
	\label{eq:A8}
\end{eqnarray}
To lowest order, $s_{2n}\sim m^n$.

\subsection{Coefficients for computing the $f_{2n}$ and $f'_{2n}$}

\begin{eqnarray}
f_0 &=& 1 \nonumber\\
f_2 &=& \frac{3}{5}\, s_2  + \frac{12}{35} \, s_2^2  + \frac{6}{175}\, s_2^3 + \frac{24}{35}\, s_2 s_4  
			+ \frac{40}{231}\, s_4^2  + \frac{216}{385}\, s_2^2 s_4  - \frac{184}{1925}\,s_2^4 
\nonumber\\
f_4 &=& \frac{1}{3} \,s_4  +\frac{18}{35}\, s_2^2  +\frac{40}{77}\, s_2 s_4  +\frac{36}{77}\, s_2^3  
			+ \frac{90}{143}\, s_2 s_6  + \frac{162}{1001}\, s_4^2 
\nonumber\\  
	&&	 \hspace{0cm}	+  ~\frac{6943}{5005}\, s_2^2 s_4  +\frac{486}{5005}\, s_2^4
\nonumber\\
f_6 &=& \frac{3}{13}\, s_6  + \frac{120}{143}\, s_2 s_4  + \frac{72}{143}\, s_2^3  + \frac{336}{715}\, s_2 s_6 
			 + \frac{80}{429}\, s_4^2 + \frac{216}{143}\, s_2^2 s_4  
\nonumber\\  
	&& 	+ ~ \frac{432}{715}\, s_2^4 
\nonumber\\
f_8 &=& \frac{3}{17}\, s_8  + \frac{168}{221}\, s_2 s_6    
			+\frac{2450}{7293}\, s_4^2  + \frac{3780}{2431}\, s_2^2 s_4  	+ \frac{1296}{2431}\, s_2^4   
\label{eq:fn}
\end{eqnarray}
\begin{eqnarray}
f'_0 &=& \frac{3}{2}  - \frac{3}{10}\, s_2^2  - \frac{2}{35}\, s_2^3 
			- \frac{1}{6}\, s_4^2 - \frac{6}{35}\, s_2^2 s_4  	+ \frac{3}{50}\, s_2^4 
\nonumber\\
f'_2 &=& \frac{3}{5}\, s_2  - \frac{3}{35}\, s_2^2  - \frac{6}{35}\, s_2 s_4  + \frac{36}{175}\, s_2^3  
			- \frac{10}{231}\, s_4^2  - \frac{17}{275}\, s_2^4 + \frac{36}{385}\, s_2^2 s_4  
\nonumber\\
f'_4 &=& \frac{1}{3}\, s_4  - \frac{9}{35}\, s_2^2  - \frac{20}{77}\, s_2 s_4  
			- \frac{45}{143}\, s_2 s_6  - \frac{81}{1001}\, s_4^2  + \frac{1}{5}\, s_2^2 s_4
\nonumber\\
f'_6 &=& \frac{3}{13}\, s_6  - \frac{75}{143}\, s_2 s_4  + \frac{270}{1001}\, s_2^3   
		- \frac{50}{429}\, s_4^2  + \frac{810}{1001}\, s_2^2 s_4  - \frac{54}{143}\, s_2^4 
\nonumber\\
	&& - ~ \frac{42}{143}\, s_2 s_6
\nonumber\\
f'_8 &=& \frac{3}{17}\, s_8  - \frac{588}{1105}\, s_2 s_6  	- \frac{1715}{7293}\, s_4^2  
		+ \frac{2352}{2431}\, s_2^2 s_4  	- \frac{4536}{12155}\, s_2^4   
\label{eq:fnp}
\end{eqnarray}

Numerical values for the $s_{2n}$ and $J_{2n}$ for a given barotrope $\rho(P)$ can be obtained through an iterative procedure. For given values of  the $J_{2n}$ and $s_{2n}$, which can initially be zero, the density distribution $\rho(l)$ is computed by numerical integration of the hydrostatic balance equation $dP/dl = -\rho(P) dU/dl$ using Eq.~(\ref{eq:A0}) and integration of the mass conservation equation $dm/dl=4\pi\,l^2\rho(l)$. Given then $\rho(l)$, new figure functions are repeatedly calculated until convergence using Eqs.~(\ref{eq:A2}) to (\ref{eq:A8}), and then the $J_{2n}$ calculated using Eq.~(\ref{eq:J2n}). Converged $J_{2n}$ values for a given barotrope require about 6 iterations of this procedure. 


\end{document}